\documentclass[twocolumn,prb,amsmath,amssymb,showpacs]{revtex4}

\usepackage{bm}% bold math

\ifx\pdftexversion\undefined
  \usepackage[dvips]{graphicx}
\else
  \usepackage[pdftex]{graphicx}
\fi

\def \be{\begin{equation}}
\def \ee{\end{equation}}
\def \bmlett{\begin{mathletters}}
\def \emlett{\end{mathletters}}

\def \k{{\bf k}}

\def \ra{\rightarrow}

\def \hx{\hat{x}}

\def \hF{\hat{F}}
\def \hI{\hat{I}}
\def \hQ{\hat{Q}}
\def \hV{\hat{V}}

\def \bS{\bar{S}}

\def \tOm{\tilde{\Omega}}

\begin{document}%******************************************************

\title{Quantum-Limited Position Detection and Amplification: A
Linear Response Perspective}
\author{A. A. Clerk}
\affiliation{Departments of Applied Physics and Physics, Yale
University, New Haven CT, 06511, USA\\
June 22, 2004}

\begin{abstract}
Using standard linear response relations, we derive the quantum
limit on the sensitivity of a {\it generic} linear-response
position detector, and the noise temperature of a generic linear
amplifier. Particular emphasis is placed on the detector's
effective temperature and damping effects; the former quantity
directly determines the dimensionless power gain of the detector.
Unlike the approach used in the seminal work of Caves [Phys. Rev.
D, {\bf 26}, 1817 (1982)], the linear-response approach directly
involves the noise properties of the detector, and allows one to
derive simple necessary and sufficient conditions for reaching the
quantum limit.  Our results have direct relevance to recent
experiments on nanoelectromechanical systems, and complement
recent theoretical studies of particular mesoscopic position
detectors.
\end{abstract}

 \maketitle

\section{Introduction}

Recent advances in the field of nanoelectromechanical (NEMS)
systems have renewed interest in the question of quantum limited
detection and amplification \cite{Blencowe, Zhang, Armour, MozTJ,
Moz, Cleland, Schwab, Bruder}; several recent experiments have
even come close to achieving this ideal limit \cite{Cleland,
Schwab}. For position measurements, the quantum limit corresponds
to the maximum sensitivity allowed by quantum mechanics of a weak,
continuous measurement \cite{Braginsky, CavesRMP}.  For
amplifiers, the quantum limit refers to the minimum amount of
noise that must be added by a high-gain linear amplifier to the
input signal \cite{Caves, DevoretNature}. Despite the fact that
quantum constraints on amplifiers have been studied and understood
for quite some time, there still seems to be some confusion in the
NEMS and mesoscopics communities as to the precise definition of
the quantum limit, and on its origin. It has even been conjectured
that it may be possible to beat the quantum limit using a weakly
coupled, continuously-measuring mesoscopic detector \cite{Zhang}.
Much of the confusion here stems from the fact that the seminal
work on quantum limited amplification by Caves \cite{Caves} uses a
description that is difficult to apply directly to the mesoscopic
detectors presently in use (i.e.~ a single-electron transistor or
quantum point contact).

In this paper, we approach the question of quantum limited
detection and amplification using nothing more than standard
linear-response (Kubo) relations.  This approach has the advantage
that it {\it directly} involves the noise properties of the
detector, and allows one to derive simple conditions that a
detector must satisfy in order to reach the quantum limit.
Achieving the quantum limit is seen to require a detector with
``ideal" noise properties, a requirement that many detectors
(e.g.~ a SET in the sequential tunnelling regime
\cite{ShnirmanSET, KorotkovSET, MakhlinRMP}) fail to meet.  We
demonstrate that the quantum limit for displacement detection and
amplification is analogous to the quantum limit constraining
quantum non-demolition (QND) measurements of a qubit
\cite{ShnirmanSET, KorotkovSET, MakhlinRMP, FirstAverin, Pilgram,
Me, Averin, ShnirmanSN}, despite the fact that in the present
problem, the detector-system coupling is not QND-- the coupling
Hamiltonian does not commute with the Hamiltonian of the input
system, and thus back-action force noise results in additional
output noise at later times. We place a special emphasis on the
effective temperature and damping effects of the detector; in
amplifier language, the latter corresponds to the amplifier's
input and output impedances. We find that {\it the detector's
effective temperature directly determines the dimensionless power
gain of the detector}, and also constrains correlations between
the detector's intrinsic output noise and back-action force noise.
The approach presented here sheds light on recent findings which
show that the effects of an out-of-equilibrium detector on an
oscillator can be described via an effective damping coefficient
and temperature \cite{MozTJ, Moz, Blencowe}; we show that this is
a generic feature of weakly-coupled linear response detectors.

Finally, turning to the specific case of position detection of an
oscillator, we find that {\it to reach the quantum limit on the
displacement sensitivity with a large power gain, the damping of
the oscillator must be predominantly independent of the detector}.
We also show that optimizing the displacement sensitivity (the
quantity measured in the experiments of Refs. \onlinecite{Cleland}
and \onlinecite{Schwab}) is {\it not} the same as minimizing the
smallest detectable force \cite{Moz}.  Note that linear-response
constraints on position measurements were also considered briefly
by Averin in Ref. \onlinecite{FirstAverin}, though that work did
not consider the role of detector-dependent damping or the
detector's effective temperature, two crucial elements of the work
presented here.

\section{Basics}

For definiteness, we start by considering the case of a generic
linear-response detector measuring the position of a harmonic
oscillator; the almost equivalent case of a generic linear
amplifier will be discussed in Section V. Our generic position
detector has an input and output port, each of which is
characterized by an operator ($\hF$ and $\hI$, respectively). The
input operator $\hF$ is linearly coupled to the position $\hx$ of
the oscillator:
\begin{equation}
    H_{int} = - A \hF \cdot \hx ,
\end{equation}
where $A$ is the dimensionless coupling strength and  $A \cdot
\hF$ is nothing more than the back-action force associated with
the measurement. The expectation value of the output operator
$\hI$ (e.g.~ current) responds to the motion of the oscillator; we
assume throughout that the coupling is weak enough that one can
use linear-response, and thus we have:
\begin{equation}
    \Delta \langle \hI(t) \rangle =
        A \int_{-\infty}^{\infty} dt' \lambda(t-t') \langle \hx(t') \rangle
\end{equation}
where $\lambda$ is the detector gain, given by the Kubo formula:
\begin{equation}
    \lambda(t) = -\frac{i}{\hbar}
         \theta(t)
            \left\langle \left[
                \hI(t), \hF(0)
                 \right] \right\rangle
    \label{gain}
\end{equation}
The expectation value here is over the (stationary) zero-coupling
density matrix $\rho_0$ of the detector.  Neither this state nor
the Hamiltonian of the detector need to be specified in what
follows.

\subsection{Effective Environment for Oscillator}

Turning to the oscillator, we assume that it is coupled both to
the detector and to an equilibrium Ohmic bath with temperature
$T_{bath}$. The bath models intrinsic (i.e.~ detector-independent)
dissipation of the oscillator.  For a weak coupling to the
detector ($A \ra 0$), one can calculate the oscillator Keldysh
Green functions using lowest-order-in-$A$ perturbation theory. One
finds that at this level of approximation, the full quantum
dynamics of the oscillator is described by a classical-looking
Langevin equation (see Appendix A and Ref.
\onlinecite{ShnirmanSN}):
\begin{eqnarray}
    m \ddot{x}(t) & = & -m \Omega^2 x(t) -
    \gamma_0 \dot{x}(t)
    - A^2 \int dt' \gamma(t-t') \dot{x}(t') \nonumber \\
    &&
    + F_{0}(t) +
    A \cdot F(t)
    \label{Langevin}
\end{eqnarray}
In this Langevin equation, $x(t)$ is a classically fluctuating
quantity, not an operator. Its average value, as determined from
Eq. (\ref{Langevin}), corresponds to the full quantum-mechanical
expectation of the operator $\hx(t)$.  Similarly, the noise in
$x(t)$ calculated from Eq. (4) corresponds precisely to $\bS_x(t)
= \langle \{\hx(t),\hx(0)\} \rangle / 2$, the symmetrized noise in
the quantum operator $\hx$ (see Appendix A for the details of this
correspondence).  Here, $\Omega$ is the renormalized frequency of
the oscillator, $m$ its renormalized mass, $\gamma_0$ describes
damping due to the equilibrium bath, and $F_0(t)$ is the
corresponding fluctuating force.  The spectrum of the $F_0$
fluctuations are given by the standard equilibrium relation:
\begin{equation}
    \bS_{F_0}(\omega) = \gamma_0 \phantom{\cdot}
        \hbar \omega \coth \left(\frac{\hbar \omega}{2 k_B T}
    \right)
    \label{EquilibSF}
\end{equation}
The remaining terms in Eq.~(\ref{Langevin}) describe the influence
of the detector-- $A \cdot F(t)$ is the random back-action force
produced by the detector, while $A^2 \gamma$ describes damping due
to the detector. The spectrum of the $F(t)$ fluctuations is given
by the symmetrized force noise of the detector:
\begin{eqnarray}
    \bS_F(\omega) & \equiv & \frac{1}{2}
        \int_{-\infty}^{\infty} dt
                e^{i \omega t}
        \langle \{ \hF(t),\hF(0) \} \rangle
    \label{SFSymm}
\end{eqnarray}
while for $\gamma(t)$, one has:
\begin{eqnarray}
    \gamma(\omega)
    & = & \frac{-\textrm{Im } \lambda_F(\omega)}{\omega}
     \equiv  \frac{ \textrm{Re }
        \int_{0}^{\infty} dt
            \left\langle \left[
                \hF(t), \hF(0)
                 \right] \right\rangle e^{i \omega t} } {\hbar \omega}
                 \nonumber \\
    & = &
     \frac{1}{\hbar}
        \frac{S_F(\omega) - S_F(-\omega)}{2 \omega}
        \label{gamma}
\end{eqnarray}
where $\lambda_F$ is the linear response susceptibility describing
the response of $F$ to a change in $x$, and $S_F(\omega)$ is the
(unsymmetrized) detector $F$-noise, calculated at zero coupling:
\begin{equation}
    S_F(\omega) = \int_{-\infty}^{\infty} dt
        \langle \hF(t) \hF(0) \rangle e^{i \omega t}.
\end{equation}
Thus, though the detector is not assumed to be in equilibrium, by
treating its coupling to the oscillator to lowest order, we obtain
a simple Langevin equation in which the detector provides both
damping and a fluctuating force.  Note that in general, the
detector force noise $\bS_F(\omega)$ will not be related to
$\gamma(\omega)$ via the temperature, as would hold for an
equilibrium system (i.e.~ Eq.~(\ref{EquilibSF})). However, {\it
for a given frequency $\omega$} we can define the effective
temperature $T_{eff}(\omega)$ via:
\begin{equation}
    \coth \left(
        \frac{\hbar \omega}{2 k_B T_{eff}(\omega)}
    \right)
        \equiv
    \frac{S_F(\omega) + S_F(-\omega)}{S_F(\omega) - S_F(-\omega)}
    \label{TEffDefn}
\end{equation}
In the $\omega \ra 0$ limit this reduces to:
\begin{equation}
    2 k_B T_{eff} = \frac{\bS_F}{\gamma} \Big|_{\omega = 0}
    \label{TeffDefn}
\end{equation}
Note that $T_{eff}$ is by no means equal to the physical
temperature of the detector, nor does it correspond to the ``noise
temperature" of the detector (see Sec. V);  the effective
temperature only serves as a measure of the asymmetry of the
detector's quantum noise.  For example, it has been found for SET
and tunnel junction detectors that $k_B T_{eff} \simeq e V$, where
$V$ is the source-drain voltage of the detector \cite{Armour,
MozTJ, Moz}.

\subsection{Detector Output Noise}

Next, we link fluctuations in the position of the oscillator to
noise in the output of the detector.  On a strictly classical
level, we would treat both the oscillator position $x(t)$ and the
detector output $I(t)$ as classically fluctuating quantities.
Using the linearity of the detector's response, we could then
write:
\begin{eqnarray}
    \delta I_{total}(\omega) & = & \delta I_0(\omega) +
    A \lambda(\omega) \cdot
    \delta x(\omega)
\end{eqnarray}
The first term ($\delta I_0$) describes the intrinsic
(oscillator-independent) fluctuations in the detector output,
while the second term corresponds to the amplified fluctuations of
the oscillator.  These are in turn given by Eq.~(\ref{Langevin}):
\begin{eqnarray}
    \delta x(\omega) & = & -\left[ \frac{1/m}{ (\omega^2 - \Omega^2) + i
    \omega \Omega / Q(\omega)} \right] (F_0(\omega) + A \cdot
    F(\omega)) \nonumber \\
        & \equiv &
         -  g(\omega) (F_0(\omega) + A \cdot
    F(\omega))
\end{eqnarray}
where $Q(\omega) = m \Omega / (\gamma_0 + \gamma(\omega))$ is the
oscillator quality factor. It follows that the total noise in the
detector output is given {\it classically} by:
\begin{eqnarray}
    S_{I,tot}(\omega) & = &
        S_{I}(\omega) +
        |g(\omega)|^2 |\lambda(\omega)|^2 \left(A^4 S_F(\omega) + A^2
            S_{F_0}(\omega) \right) \nonumber \\
        &&         \label{SItot}
        - 2 A^2 \textrm{Re }
            \left[ g(\omega) S_{I F}(\omega) \right]
\end{eqnarray}
Here, $S_I, S_F$ and $S_{IF}$ are the (classical) detector noise
correlators calculated in the absence of any coupling to the
oscillator.

To apply the classically-derived Eq.~(\ref{SItot}) to our quantum
detector-plus-oscillator system, we interpret $S_{I,tot}$ as the
total symmetrized quantum-mechanical output noise of the detector,
and simply substitute in the right-hand side the symmetrized
quantum-mechanical detector noise correlators $\bS_F$, $\bS_I$ and
$\bS_{IF}$, defined as in Eq.~(\ref{SFSymm}). Though this may seem
rather ad-hoc, one can easily demonstrate that Eq.~(\ref{SItot})
thus interpreted would be completely rigorous,
quantum-mechanically, {\it if} the detector correlation functions
obeyed Wick's theorem. Thus, quantum corrections to
Eq.~(\ref{SItot}) will arise from the non-Gaussian nature of the
detector noise correlators.  We expect from the central limit
theorem that such corrections will be small in the relevant limit
where $\omega$ is much smaller that the typical detector frequency
$\sim k_B T_{eff} / \hbar$, and neglect these corrections in what
follows.  Note that the validity of Eq.~(\ref{SItot}) for a
specific model of a tunnel junction position detector has been
verified in Ref. \onlinecite{CGTunnel}.

\subsection{Quantum Constraint on Detector Noise}

We turn now to the fundamental constraint on the detector noise
correlators and gain $\lambda$ which, in the present approach,
serves as the basis of the quantum limit. We have:
\begin{equation}
    \bS_I(\omega) \bS_F(\omega) \geq \frac{\hbar^2}{4}
        \left[ \textrm{Re } (\lambda(\omega) - \lambda'(\omega))
        \right]^2 +
        \left[\textrm{Re } \bS_{IF}(\omega) \right]^2
        \label{NoiseConstraint}
\end{equation}
Here, $\lambda'$ is the reverse gain of the detector (i.e.~ the
gain in an experiment where we couple $x$ to $I$ and attempt to
measure $F$).  Eq.~(\ref{NoiseConstraint}) tells us that if our
detector has gain and no positive feedback ($\textrm{Re }\lambda
\cdot \textrm{Re }\lambda' \leq 0$), then it must have a minimum
amount of back-action and output noise.  This equation was proved
rigorously in Ref. \onlinecite{Me} for $\omega=0$ using a Schwartz
inequality; the proof given there may be straightforwardly
generalized to finite $\omega$ if one now uses
symmetrized-in-frequency noise correlators.  Note that almost all
mesoscopic detectors that have been studied in detail (i.e.~ a SET
or generalized quantum point contact) have been found to have
$\lambda' = 0$ \cite{Me}.

We now define a quantum-limited detector at frequency $\omega$ as
having a minimal amount of noise at $\omega$, that is, it
satisfies:
\begin{equation}
    \bS_I(\omega) \bS_F(\omega) = \frac{\hbar^2}{4}
        \left[ \textrm{Re } (\lambda(\omega))
        \right]^2 +
        \left[\textrm{Re } \bS_{IF}(\omega) \right]^2.
        \label{CorrQLCondition}
\end{equation}
This is the same condition that arose in the study of
quantum-limited qubit detection \cite{Me, Averin}, with the
exception that in that case, one also required that $\textrm{Re }
\bS_{IF} = 0$.  We will show that Eq.~(\ref{CorrQLCondition}) must
indeed be satisfied in order to achieve the quantum limit on
position sensitivity, or the quantum limit on the noise
temperature of an amplifier.

As discussed in Ref. \onlinecite{Me}, having a quantum-limited
detector (i.e.~ satisfying Eq.~(\ref{CorrQLCondition})) implies a
tight connection between the input and output ports of the
detector.  Note first that we may write each symmetrized noise
correlators as a sum over transitions $|i \rangle \ra | f
\rangle$, e.g.~
\begin{eqnarray}
    \bS_{F}(\omega) & = & \pi \hbar \sum_{i,f}
        \langle i | \rho_0 | i \rangle
        | \langle f | F | i \rangle |^2
         \label{LehmanRep} \\
        && \times
        \left[
            \delta(E_f-E_i + \omega) +
            \delta(E_f-E_i - \omega)
        \right], \nonumber
\end{eqnarray}
where $\rho_0$ is the stationary detector density matrix, and $|
i/f \rangle$ is a detector eigenstate with energy $E_{i/f}$.  To
achieve the ``ideal noise" condition of
Eq.~(\ref{CorrQLCondition}) at frequency $\omega$, there must
exist a complex factor $\alpha$ (having dimensions $[I] / [F]$)
such that:
\begin{equation}
    \langle f | I | i \rangle = \alpha \langle f | F | i \rangle
    \label{PropCond}
\end{equation}
for {\it each} pair of initial and final states $|i \rangle$, $|f
\rangle$ contributing to $\bS_F(\omega)$ and $\bS_{I}(\omega)$
(c.f.~ Eq.~(\ref{LehmanRep})).  Note that this {\it not} the same
as requiring Eq.~(\ref{PropCond}) to hold for all possible states
$|i\rangle$ and $| f \rangle$. In Ref. \onlinecite{Me}, the
requirement of Eq.~(\ref{PropCond}) was further interpreted as
implying that there is no additional information regarding the
input signal that is available in the detector but not revealed in
its output $\hI$.

For a quantum limited detector with $\lambda'=0$, the coefficient
$\alpha$ in Eq.~(\ref{PropCond}) can be found from:
\begin{eqnarray}
    | \alpha(\omega) |^2 & = &  \bS_I(\omega) / \bS_F(\omega)  \\
    \tan \left( \arg \alpha (\omega) \right) & = &
        -\frac{ \hbar \textrm{Re } \lambda(\omega)/2}
         {\textrm{Re } \bS_{IF}(\omega)}
\end{eqnarray}
Thus, a non-vanishing gain implies that $\textrm{Im } \alpha \neq
0$.  It then follows from Eq.~(\ref{PropCond}) and the hermiticity
of $\hI$,$\hF$ that for a quantum limited detector, the set of all
initial states $|i\rangle$ contributing to the noise has no
overlap with the set of all final states $| f \rangle$.  This
immediately implies that a quantum limited detector cannot be in
equilibrium.

\subsection{Power Gain}

To be able to say that our detector truly amplifies the motion of
the oscillator, the power delivered by the detector to a following
amplifier must be much larger than the power used to drive the
oscillator-- i.e.~, the detector must have a dimensionless power
gain $G_P(\omega)$ much larger than one.  As we now show, this
requirement places further constraints on the effective
temperature and noise properties of the detector.

In what follows, we consider the simple (and usual) case where
there is no reverse gain: $\lambda' = 0$. The power gain
$G_P(\omega)$ of our generic position detector may be defined as
follows.  Imagine first that we drive the oscillator with a force
$F_{D} \cos \omega t$; this will cause the output of our detector
$\langle I(t) \rangle$ to also oscillate at frequency $\omega$. To
optimally detect this signal in the detector output, we further
couple $I$ to a second oscillator with natural frequency $\omega$
and position $y$: $H'_{int} = B \hI \cdot \hat{y}$.  The
oscillations in $I$ will now act as a driving force on the
auxiliary oscillator $y$. $G_P(\omega)$ is then defined as the
{\it maximum} ratio between the power provided to the output
oscillator $y$ from the detector, versus the power fed into the
input of the amplifier. This ratio is maximized if there is no
intrinsic (i.e.~ detector-independent) damping of either the input
or output oscillators.  The damping of of the input oscillator is
then completely given by $A^2 \gamma$ (c.f.~ Eq.~(\ref{gamma})),
whereas the the damping of the output oscillator $B^2
\gamma_{out}$ is given by:
\begin{eqnarray}
    \gamma_{out}(\omega)
    & = & -\frac{\textrm{Im } \lambda_I(\omega)}{\omega}
    \equiv \frac{ \textrm{Re }
        \int_0^{\infty} dt
            \left\langle \left[
                \hI(t), \hI(0)
                 \right] \right \rangle e^{i \omega t}} {\hbar \omega}
                 \nonumber \\
    \label{gammaout}
\end{eqnarray}
Using a bar to denote a time average, we then have:
\begin{eqnarray}
    G_P(\omega)
        & \equiv &
            \frac{\overline{P_{out}}}{\overline{P_{in}}}
            =
            \frac{B^2 \gamma_{out}(\omega) \cdot \overline{ \dot{y}^2} }
            {A^2 \gamma(\omega) \cdot \overline{ \dot{x}^2 } } \\
        & = &
            \frac{B^2 \gamma_{out}(\omega)}{A^2 \gamma(\omega)}
                \times
                \nonumber \\
        && \frac{
            \left[
                \left(
                    \frac{\omega}{\omega B^2 \gamma_{out}(\omega)}
                \right)
                \left(
                    B \cdot A |\lambda(\omega)| \cdot
                    |g(\omega)| F_{ext}
                 \right) \right]^2}
            {
                \left[ \omega |g(\omega)| F_{ext} \right]^2}
            \nonumber \\
        & = &
            \frac{ |\lambda(\omega)|^2 }
                {\omega^2 \gamma_{out}(\omega) \cdot \gamma(\omega) }
               \nonumber \\
        & = &
            \frac{ |\lambda(\omega)|^2}
                {\textrm{Im } \lambda_F(\omega)
                \cdot \textrm{Im } \lambda_I(\omega)}
                \label{GPDefn}
\end{eqnarray}
 Thus, the power gain is a simple dimensionless ratio formed by
 the three different response coefficients characterizing the detector.
It is completely analogous to the power gain of a voltage
amplifier (see Eq.~(\ref{GPVoltAmp}) of Sec. V)

Turning now to the important case of a quantum limited detector,
that is a detector satisfying the ideal noise condition of
Eq.~(\ref{CorrQLCondition}), we find that the expression for the
power gain can be further simplified using Eq.~(\ref{PropCond}).
One finds:
\begin{equation}
   G_P(\omega) =
       \frac{ \left(\textrm{Im } \alpha \right)^2
                   \coth \left( \frac{\hbar \omega}{2 k_B T_{eff}}
                   \right)
       +   \left( \textrm{Re } \alpha \right)^2 }
       {|\alpha|^2 / 4}
       \label{GPTemp}
\end{equation}
It thus follows that to have $G_P \gg 1$, one needs $\k_B T_{eff}
\gg \hbar \omega$: {\it a large power gain implies a large
effective detector temperature}.  In the large $G_P$ limit, we
have
\begin{equation}
    G_P \simeq \left[ \frac{\textrm{Im } \alpha }{ |\alpha| }
        \frac{4 k_B T_{eff}}{\hbar \omega }\right]^2
\end{equation}

Finally, an additional consequence of the large $G_P(\omega)$,
large $T_{eff}$ limit is that the imaginary parts of the gain
$\lambda(\omega)$ and cross-correlator $\bS_{IF}(\omega)$ become
negligible; they are suppressed by the small factor $\hbar \omega
/ k_B T_{eff}$. This is shown explicitly in Appendix B.

\section{Quantum Limit on Added Displacement Noise}

The sensitivity of a position detector is determined by the added
displacement noise, $S_{x}(\omega)$, which is simply the total
detector contribution to the noise in the detector's output,
referred back to the oscillator.  It is this quantity which has
been probed in recent experiments \cite{Cleland,Schwab}, and which
has a fundamental quantum constraint, as we derive below.

To define $S_{x}(\omega)$, we first introduce $S_{x,tot}(\omega)$,
which is simply the total noise in the output of the detector
($S_{I,tot}(\omega)$, c.f.~ Eq.~(\ref{SItot})) referred back to
the oscillator:
\begin{equation}
    S_{x,tot}(\omega) \equiv \frac{S_{I,tot}(\omega)}{A^2 \lambda^2}
\end{equation}
We then separate out detector-dependent contributions to
$S_{x,tot}(\omega)$ from the intrinsic equilibrium fluctuations of
the oscillator; the added displacement noise $S_x(\omega)$ is
defined as the former quantity:
\begin{equation}
    S_{x,tot}(\omega) \equiv S_x(\omega)  +
            \frac{\gamma_0}{\gamma_0 + A^2 \gamma} S_{x,eq}(\omega,T)
    \label{SxTot}
\end{equation}
where:
\begin{equation}
    S_x(\omega)   =
        \frac{\bS_I}{|\lambda|^2 A^2} + A^2 |g(\omega)|^2 \bS_F -
        \frac{
        2 \textrm{Re }
            \left[ \lambda^* g^*(\omega)
            \bS_{IF} \right]}
            { |\lambda|^2}
            \label{Sx}
\end{equation}
\vspace{-0.8 cm}
\begin{equation}
    S_{x,eq}(\omega,T)   =
        \hbar
        \coth\left( \frac{\hbar \omega}{2 k_B T} \right)
        \left[ -\textrm{Im } g(\omega) \right]
        \label{SxEquilib}
\end{equation}
and where we have omitted writing the $\omega$ dependence of the
noise correlators and $\lambda$.  Here, $S_{x,eq}(\omega,T)$
represents the equilibrium fluctuations that would result if {\it
all} the damping were due to the equilibrium bath; its
contribution to $S_{x,tot}(\omega)$ is reduced, as only part of
the damping results from the bath.

We now proceed to derive the quantum limit on $S_x(\omega)$.
Examining Eq.~(\ref{Sx}) for $S_x(\omega)$, and ignoring for a
moment the detector-dependent damping of the oscillator, we see
that the first term (i.e.~ the intrinsic detector output noise
referred back to the detector input) is proportional to $1/A^2$,
while the second term (i.e.~ the back-action of the detector)
scales as $A^2$. We would thus expect $S_x(\omega)$ to attain a
minimum value at an optimal choice of coupling $A = A_{opt}$ where
both these terms make equal contributions.  Using the inequality
$X^2 + Y^2 \geq 2 X Y$ we see that this value serves as a lower
bound on $S_x$ even in the presence of detector-dependent damping.
Defining $\phi = \arg g(\omega)$, we thus have the bound:
\begin{equation}
        S_x(\omega)  \geq  2 | g(\omega) | \left[
            \sqrt{ \bS_I \bS_F / |\lambda|^2 }
            -   \frac{
                \textrm{Re }
                    \left[
                        \lambda^* e^{-i \phi(\omega)} \bS_{IF}
                    \right]}
                {|\lambda|^2}
        \right]
        \label{FirstMin}
\end{equation}
where the minimum value is achieved when:
\begin{equation}
    A^2_{opt} = \sqrt{ \frac{\bS_I(\omega) }
        {|\lambda(\omega) g(\omega)|^2 \bS_F(\omega)}}
    \label{AOptOffRes}
\end{equation}
In the case where the detector-dependent damping is negligible,
the RHS of this equation is independent of $A$, and thus
Eq.~(\ref{AOptOffRes}) can be satisfied by simply tuning the
coupling strength $A$; in the more general case where there is
detector-dependent damping, the RHS is also a function of $A$, and
it may no longer be possible to achieve Eq.~(\ref{AOptOffRes}) by
simply tuning $A$.

While Eq.~(\ref{FirstMin}) is certainly a bound on the added
displacement noise $S_x(\omega)$, it does not in itself represent
the quantum limit.  Reaching the quantum limit requires more than
simply balancing the detector back-action and intrinsic output
noises (i.e.~ the first two terms in Eq.~(\ref{Sx})); {\it one
also needs a detector with ideal noise properties, that is a
detector which satisfies Eq.~(\ref{CorrQLCondition}) and thus the
proportionality condition of Eq.~(\ref{PropCond})}. Using the
quantum noise constraint of Eq.~(\ref{NoiseConstraint}) to further
bound $S_x(\omega)$, we obtain:
\begin{eqnarray}
        S_x(\omega)  & \geq&
            2 \frac{| g(\omega) |}{|\lambda|} \Bigg[
            \sqrt{
                \left(\frac{\hbar \textrm{Re }\lambda}{2} \right)^2 +
                \left( \textrm{Re } \bS_{IF} \right)^2}
            \nonumber \\
            && -   \frac{
                \textrm{Re }
                    \left[
                        \lambda^* e^{-i \phi(\omega)} S_{IF}
                    \right]}
                {|\lambda|^2}
        \Bigg]
        \label{SecondMin}
\end{eqnarray}
The minimum value of $S_x(\omega)$ in Eq.~(\ref{SecondMin}) is now
achieved when one has {\it both} an optimal coupling (i.e.~
Eq.~(\ref{AOptOffRes})) {\it and} a quantum limited detector, that
is one which satisfies Eq.~(\ref{CorrQLCondition}).  Note again
that {\it an arbitrary detector will not satisfy the ideal noise
condition of Eq.~(\ref{CorrQLCondition})}.

Next, we consider the relevant case where our detector is a good
amplifier and has a power gain $G_P(\omega) \gg 1$ over the width
of the oscillator resonance. As discussed in Appendix B, this
implies that we may neglect the imaginary parts of $\lambda$ and
$\bS_{IF}$, as they are suppressed by $\hbar \Omega / k_B T_{eff}
\ll 1$. We then have:
\begin{equation}
        S_x(\omega)  \geq  2 | g(\omega) | \left[
            \sqrt{
                \left(\frac{\hbar }{2} \right)^2 +
                \left(  \frac{\bS_{IF}}{ \lambda }\right)^2}
            -   \frac{
                \cos \left[\phi(\omega)\right] \bS_{IF}
                    }
                {\lambda}
        \right]
        \label{ThirdMin}
\end{equation}
Finally, as there is no further constraint on $ \bS_{IF} /
\lambda$, we can minimize the expression over its value.  The
minimum $S_{x}(\omega)$ is achieved for a detector whose
cross-correlator satisfies:
\begin{equation}
    \frac{\bS_{IF}(\omega)}{\lambda} \Big|_{optimal} =
        \frac{\hbar}{2} \cot \phi(\omega),
        \label{OptSIF}
\end{equation}
with the minimum value being given by:
\begin{equation}
S_x(\omega) \Big|_{min} = \hbar | \textrm{Im }g(\omega) | =
    \lim_{T\ra0} S_{x,eq}(\omega,T)
    \label{CCBound}
\end{equation}
where $S_{x,eq}(\omega,T)$ is the equilibrium contribution to
$S_{x,tot}(\omega)$ defined in Eq.~(\ref{SxEquilib}).  Thus, in
the limit of a large power gain, we have that {\it at each
frequency, the minimum displacement noise due to the detector is
precisely equal to the noise arising from a zero temperature
bath}.  This conclusion is irrespective of the strength of the
intrinsic (detector-independent) oscillator damping.

The result of Eq.~(\ref{CCBound}) is essentially identical to the
conclusion of Caves \cite{Caves}, who found that a high-gain
amplifier (modelled as a set of bosonic modes and a scattering
matrix) must add at least $\hbar \omega / 2$ of noise to an input
signal at frequency $\omega$. Here, our input signal corresponds
to the damped oscillator, and the minimum value of $S_x(\omega)$
in Eq.~(\ref{CCBound}) corresponds precisely to the zero-point
noise of the damped oscillator.

Though it reaches a similar conclusion, the linear-response
approach has several advantages over the approach of Caves.
First, we do not have to model our detector as a set of bosonic
modes and a scattering matrix, something that is difficult to do
for many mesoscopic detectors. More significantly, the
linear-response approach makes explicitly clear what is needed to
reach the quantum limit.  We find that to reach the quantum-limit
on the added displacement noise $S_x(\omega)$, one needs:
\begin{enumerate}
\item A quantum limited detector, that is a detector which
satisfies the ``ideal noise" condition of
Eq.~(\ref{CorrQLCondition}), and hence the proportionality
condition of Eq. (\ref{PropCond}).

\item A coupling $A$ which satisfies Eq.~(\ref{AOptOffRes}).

\item A detector cross-correlator $\bS_{IF}$ which satisfies
Eq.~(\ref{OptSIF}).
\end{enumerate}
Note that condition (i) is identical to what is required for
quantum-limited detection of a qubit; it is rather demanding, and
requires that there is no ``wasted" information about the input
signal in the detector which is not revealed in the output
\cite{Me}. Also note that $\cot \phi$ changes quickly as a
function of frequency across the oscillator resonance, whereas
$\bS_{IF}$ will be roughly constant; condition (ii) thus implies
that it will not be possible to achieve a minimal $S_x(\omega)$
across the entire oscillator resonance.  A more reasonable goal is
to optimize $S_x$ at resonance, $\omega = \Omega$.  As $g(\Omega)$
is imaginary, Eq.~(\ref{OptSIF}) tells us that $\bS_{IF}$ should
be zero. Assuming we have a quantum-limited detector with a large
power gain ($k_B T_{eff} \gg \hbar \Omega$), the remaining
condition on the coupling $A$ (Eq.~(\ref{AOptOffRes})) may be
written as:
\begin{equation}
    \frac{A_{opt}^2 \gamma}{\gamma_0 + A_{opt}^2 \gamma} =
    \left| \frac{\textrm{Im } \alpha}{\alpha}  \right|
    \frac{1}{\sqrt{G_P(\Omega)}}
    = \frac{\hbar \Omega}{4 k_B T_{eff}}
    \label{OptA2}
\end{equation}
As $A^2 \gamma$ is the detector-dependent damping of the
oscillator, we thus have that {\it to achieve the quantum-limited
value of $S_x(\Omega)$ with a large power gain, one needs the
intrinsic damping of the oscillator to be much larger than the
detector-dependent damping}.  The detector-dependent damping must
be small enough to compensate the large effective temperature of
the detector; if the bath temperature satisfies $\hbar \Omega /
k_B \ll T_{bath} \ll T_{eff}$, Eq.~(\ref{OptA2}) implies that at
the quantum limit, the temperature of the oscillator will be given
by:
\begin{equation}
    T_{osc} \equiv \frac{ A^2 \gamma \cdot T_{eff} +
        \gamma_0 \cdot T_{bath}}
    { A^2 \gamma  + \gamma_0}
    \ra
    \frac{\hbar \Omega}{4 k_B} + T_{bath}
\end{equation}
Thus, at the quantum limit and for large $T_{eff}$, the detector
raises the oscillator's temperature by $\hbar \Omega / 4 k_B$.  As
expected, this additional heating is only {\it half} the zero
point energy; in contrast, the quantum-limited value of
$S_x(\omega)$ corresponds to the full zero point result, as it
also includes the contribution of the intrinsic output noise of
the detector.

Finally, we remark that if one did not assume $k_B T_{eff} \gg
\hbar \Omega$ as is needed for a large power gain, we would have
to keep the imaginary parts of $\lambda$ and $\bS_{IF}$.  One can
show that for $k_B T_{eff} / \hbar \Omega \ra 0$, it is possible
to have a perfect anti-correlation between the intrinsic detector
output noise $\delta I_0$ and the back-action force $\delta F$,
and thus have $S_{x}(\omega) = 0$. Thus, similar to the results of
Caves \cite{Caves}, in the limit of unit power gain (i.e.~ small
detector effective temperature), there is no quantum limit on
$S_x$, as perfect anti-correlations between the two kinds of
detector noise (i.e.~ in $\hI$ and $\hF$) are possible.

\section{Quantum Limit on Force Sensitivity}

In this section, we now ask a different question: what is the
smallest magnitude force acting on the oscillator that can be
detected with our displacement detector \cite{Moz}?  This force
sensitivity is also subject to a quantum limit; as with the
quantum limit on $S_x(\omega)$, one again needs a quantum-limited
detector (i.e.~ one satisfying the ideal noise condition of
Eq.~(\ref{CorrQLCondition})) to reach the maximal sensitivity.
However, the conditions on the coupling $A$ (i.e.~ the
detector-dependent damping) are quite different than what is
needed to optimize $S_x(\omega = \Omega)$; in particular, one can
reach the quantum limit on the force sensitivity even if there is
no intrinsic oscillator damping.

We start by imagining that a force $F_{ext}(t) = \Delta p \cdot
\delta(t)$ acts on our oscillator, producing a change $\Delta
I(t)$ in the output of our detector .  The corresponding
signal-to-noise ratio is defined as \cite{Braginsky, Moz}:
\begin{equation}
    S/N = \int \frac{d \omega}{2 \pi} \frac{ | \Delta I(\omega)
    |^2 }{S_{I,tot}(\omega)} = (\Delta p)^2
    \int \frac{d \omega}{2 \pi} \frac{ | \lambda(\omega) g(\omega)
      |^2 }{S_{I,tot}(\omega)}
    \label{SNDefn}
\end{equation}
In what follows, we make the reasonable assumption that the
detector noise correlators and gain are frequency independent over
the width of the oscillator resonance.  We also take the relevant
limit of a large power gain (i.e.~ $k_B T_{eff} \gg \hbar
\Omega$), and assume a best-case scenario where $T_{bath} = 0$ and
where the total oscillator quality factor $Q \gg 1$. Using
Eq.~(\ref{SItot}), one finds that the maximal $S/N$ is indeed
obtained for a quantum-limited detector (i.e.~ one satisfying
Eq.~(\ref{CorrQLCondition})):
\begin{equation}
    S/N \leq \frac{ 2 (\Delta p)^2}{\hbar m \tOm}
        \int^{\infty}_0 \frac{d x}{\pi}
            \frac{\Gamma}{(x^2-1)^2 + \Gamma^2(1 + \Lambda^2 x)}
            \label{MinSN1}
\end{equation}
where equality occurs for a quantum-limited detector.  For such a
detector, we have:
\begin{eqnarray}
    \tOm^2 & = &
        \Omega^2 - \left(\frac{1}{Q_{det}}\right)
                \left(\frac{4 k_B T_{eff}}{\hbar
        \Omega}\right)
        \left( \frac{\textrm{Im }\alpha \textrm{Re }
        \alpha}{|\alpha|^2}\right) \\
    \Gamma & = &
        \left(\frac{1}{Q_{det}}\right)
        \left( \frac{4 k_B T_{eff}}{\hbar
        \Omega}\right) \left(\frac{\textrm{Im }\alpha}{|\alpha|}
        \frac{\Omega}{\tOm} \right)^2  \\
    \Lambda^2 & = &
        \left(\frac{\gamma_0}{A^2 \gamma}\right)
        \left( \frac{\hbar \tOm}{4 k_B
        T_{eff}} \right)
        \left( \frac{|\alpha|}{\textrm{Im }\alpha} \right)
\end{eqnarray}
where $Q_{det} = m \Omega / (A^2 \gamma)$ is the quality factor
corresponding to the detector-dependent damping.  It is clear from
Eq.~(\ref{MinSN1}) that the signal to noise ratio can be further
maximized by having both $\Gamma \ll 1$ and $\Lambda \ll 1$.  This
requires:
\begin{equation}
    \frac{\gamma_0}{A^2 \gamma} \ll
    \frac{k_B T_{eff}}{\hbar \Omega}  \ll  Q_{det}
    \label{SNOptA}
\end{equation}
If this condition is satisfied, it follows that $\tOm \simeq
\Omega$, and we have:
\begin{equation}
    S/N \leq \frac{(\Delta p)^2}{\hbar M \Omega}
\end{equation}
Demanding now $S/N \geq 1$, we find that the minimum detectable
$\Delta p$ is $\sqrt{\hbar M \Omega}$, $\sqrt{2}$ times the zero
point value. Note that the requirement of Eq.~(\ref{SNOptA}) on
the coupling $A$ is very different from what is needed to reach
the quantum limit on $S_x(\Omega)$ (c.f.~ Eq.~(\ref{OptA2})).  In
the present case, it is possible to reach the quantum limit on the
force sensitivity even if the damping from the detector dominates
($A^2 \gamma \gg \gamma_0$); in contrast, it is impossible to
achieve the quantum limit on $S_x(\Omega)$ in this regime.

\section{Quantum Limit on the Noise Temperature of a Voltage Amplifier}

In this final section, we generalize the discussion of the
previous sections to the case of a generic linear voltage
amplifier (see, e.g.~,Ref. \onlinecite{DevoretNature}); the
quantum limit on the noise temperature of the amplifier is seen to
be analogous to the quantum limit on the added displacement noise
$S_x(\omega)$.

As with the position detector, the voltage amplifier is
characterized by an input operator $\hQ$ and an output operator
$\hV$; these play the role, respectively, of $\hF$ and $\hI$ in
the position detector.  $\hV$ represents the output voltage of the
amplifier, while $\hQ$ is the operator which couples to the input
signal $v_{in}(t)$ via a coupling Hamiltonian:
\begin{equation}
    H_{int} = v_{in}(t) \cdot \hQ
\end{equation}
In more familiar terms,  $\tilde{I}_{in} = -d \hQ / dt$ represents
the current flowing into the amplifier.  We also assume that the
output of the amplifier ($\hV$) is connected to an external
circuit via a term:
\begin{equation}
    H'_{int} = q_{out}(t) \cdot \hV
\end{equation}
where $\tilde{i}_{out} = d q_{out} / dt$ is the current in the
external circuit. In what follows, we will use quantities defined
in the previous sections for a position detector, simply
substituting in $\hI \ra \hV$ and $\hF \ra \hQ$.

Similar to the position detector, there are three response
coefficients for the amplifier: the voltage gain coefficient
$\lambda$ (c.f.~ Eq.~(\ref{gain})), the $Q-Q$ susceptibility
$\lambda_Q$ which determines damping at the input (c.f.~
Eq.~(\ref{gamma})), and the $V-V$ susceptibility $\lambda_V$ which
determines damping at the output (c.f.~ Eq.~(\ref{gammaout})). The
diagonal susceptibilities determine the input and output
impedances:
\begin{eqnarray}
    Z_{in}(\omega) & = &
        \left[i \omega \lambda_Q(\omega)\right]^{-1} \\
    Z_{out}(\omega) & = &
        \frac{\lambda_V(\omega)}{-i \omega}
\end{eqnarray}
i.e.~ $\langle \tilde{I}_{in} \rangle_{\omega} =
\frac{1}{Z_{in}(\omega)} v_{in}(\omega)$ and $ \langle V
\rangle_{\omega} = Z_{out}(\omega) \tilde{i}_{out}(\omega)$, where
the subscript $\omega$ indicates the Fourier transform of a
time-dependent expectation value.

We will consider throughout this section the case of no reverse
gain, $\lambda' = 0$.   We may then use Eq.~(\ref{GPDefn}) for the
power gain; it has the expected form:
\begin{equation}
    G_P = \lambda^2 \frac{\textrm{Re }Z_{in}}
        {\textrm{Re }Z_{out}}
        = \frac{ \langle V \rangle^2 / \textrm{Re } Z_{out}}
        { (v_{in})^2 / \textrm{Re } Z_{in}}
    \label{GPVoltAmp}
\end{equation}
Finally, we may again define the effective temperature
$T_{eff}(\omega)$ of the amplifier via Eq.~(\ref{TEffDefn}), and
define a quantum-limited amplifier as one which satisfies the
ideal noise condition of Eq.~(\ref{CorrQLCondition}).  For such an
amplifier, the power gain will again be determined by the
effective temperature via Eq.~(\ref{GPTemp}).

Turning to the noise, we introduce $S_{v,tot}(\omega)$, the total
noise at the output port of the amplifier referred back to the
input. Assuming the voltage source producing the input signal
$v_{in}$ has an impedance $Z_S$ and a temperature $T_S \gg \hbar
\omega / k_B$, we may write:
\begin{equation}
    S_{v,tot}(\omega) = 2 k_B T_{s} \textrm{Re } Z_S(\omega)
        + S_{v}(\omega)
        \label{Svtot}
\end{equation}
The first term is the equilibrium noise associated with the signal
source, while the second term is the total amplifier contribution
to the noise at the output port, referred back to the input.
Taking the limit of a large power gain (which ensures $\lambda,
\bS_{VQ}$ are real), we have (c.f.~ Eq.~(\ref{Sx}) and Ref.
\onlinecite{DevoretNature}):
\begin{eqnarray}
    S_{v}(\omega) & = & \frac{\bS_V}{\lambda^2} + |\tilde{Z}|^2 \left[
        \omega^2 \bS_Q\right]
        + 2 \textrm{ Im} \tilde{Z}  \frac{\left[w \bS_{V
        Q}\right] }{\lambda}
        \label{Sv} \\
        \tilde{Z} & = &
            \frac{ Z_S Z_{in}}{Z_S + Z_{in}}
\end{eqnarray}
where for clarity, we have dropped the $\omega$ dependence of the
noise correlators, gain, and impedances.  The first term in
Eq.~(\ref{Sv}) represents the intrinsic output noise of the
amplifier, while the second term represents the back-action of the
amplifier: fluctuations in the input current $\hat{I}_{in} = -d
\hQ / dt$ lead to fluctuations in the voltage drop across the
parallel combination of the source impedance $Z_s$ and the input
impedance $Z_{in}$, and hence in the signal going into the
amplifier.  The last term in Eq.~(\ref{Sv}) represents
correlations between these two sources of noise.  We see that the
general form of Eq.~(\ref{Sv}) is completely analogous to that for
the added displacement noise $S_x(\omega)$ of a displacement
detector, c.f.~ Eq.~(\ref{Sx}).

We are now ready to introduce the noise temperature of our
amplifier: $T_N$ is defined as the amount we must increase the
temperature $T_S$ of the voltage source to account for the noise
added by the amplifier \cite{TNNote}, i.e.~ we wish to re-write
Eq.~(\ref{Svtot}) as:
\begin{equation}
    S_{v,tot}(\omega) \equiv 2 k_B (T_S + T_N) \textrm{Re } Z_s(\omega)
\end{equation}

In what follows, we assume that $|Z_s| \ll |Z_{in}|$, which means
that $Z_{in}$ drops out of Eq.~(\ref{Sv}); we will test the
validity of this assumption at the end. Using Eq.~(\ref{Sv}), and
writing $Z_s = |Z_s| e^{i \phi}$, we have immediately:
\begin{equation}
    2 k_B T_N =
        \frac{1}{\cos \phi} \left[
            \frac{\bS_V}{|Z_s| \lambda^2} + |Z_{s}| \left(
            \omega^2 \bS_Q\right) \right]
            + 2 \tan \phi   \frac{\omega \bS_{V
            Q}}{\lambda}
        \label{TNEqn}
\end{equation}

To derive a bound on $T_N$, we first perform the classical step of
optimizing over the the magnitude and phase of the source
impedance $Z_S(\omega)$; this is in contrast to the optimization
of $S_x(\omega)$, where one would optimize over the strength of
the coupling.  We find:
\begin{equation}
    k_B T_N \geq \omega \sqrt{
        \frac{
        \bS_V(\omega) \bS_Q(\omega) -
            \left[\bS_{V Q}(\omega)\right]^2 }
            {\left[\lambda(\omega)\right]^2} }
        \label{TNMin1}
\end{equation}
where the minimum is achieved for an optimal source impedance
satisfying:
\begin{eqnarray}
    |Z_s(\omega)|_{opt} & = & \sqrt{ \frac{\bS_V(\omega) /
        \left[ \lambda(\omega) \right]^2}
        {\omega^2
    \bS_Q(\omega)}}
    \label{ZOpt1} \\
    \sin \phi(\omega) \big|_{opt} & = & - \frac{\bS_{V Q}(\omega)}
        {\sqrt{\bS_V(\omega) \bS_Q(\omega)}}
        \label{ZOpt2}
\end{eqnarray}

As with the displacement sensitivity, simply performing a
classical optimization (here, over the choice of source impedance)
is not enough to reach the quantum limit on $T_N$.  One also needs
to have an amplifier which satisfies the ideal noise condition of
Eq.~(\ref{CorrQLCondition}).  Using this equation, we obtain the
final bound:
\begin{equation}
    k_B T_N \geq \frac{\hbar \omega}{2}
    \label{TNQL}
\end{equation}
where the minimum corresponds to both having optimized $Z_s$ and
having an amplifier whose noise satisfies
Eq.~(\ref{CorrQLCondition}).

Finally, we need to test our initial assumption that $|Z_S| \ll
|Z_{in}|$.  Using the proportionality condition of
Eq.~(\ref{PropCond}) and the fact that we are in the large power
gain limit ($G_P(\omega) \gg 1$), we find:
\begin{equation}
    \left|
        \frac{Z_S(\omega)}{ \textrm{Re } Z_{in} (\omega) }
    \right |  =
    \left | \frac{  \alpha }{\textrm{Im } \alpha} \right |
    \frac{\hbar \omega}{4 k_B T_{eff}} =
    \frac{1}{\sqrt{G_P(\omega)} } \ll 1
    \label{ZinConstraint}
\end{equation}
It follows that $|Z_S| \ll |Z_{in}|$ in the large power gain,
large effective temperature regime of interest, and our neglect of
$|Z_{in}|$ in Eq. (\ref{TNEqn}) is justified.  Eq.
(\ref{ZinConstraint}) is analogous to the case of the displacement
detector, where we found that reaching the quantum limit on
resonance required the detector-dependent damping to be much
weaker than the intrinsic damping of the oscillator (c.f. Eq.
(\ref{OptA2})).

Thus, similar to the situation of the displacement detector, the
linear response approach allows us both to derive rigorously the
quantum limit on the noise temperature $T_N$ of an amplifier, and
to state conditions that must be met to reach this limit.  To
reach the quantum-limited value of $T_N$ with a large power gain,
one needs {\it both} a tuned source impedance $Z_S$, {\it and} an
amplifier which possesses ideal noise properties (c.f.~
Eq.~(\ref{CorrQLCondition}) and Eq. (\ref{PropCond})).

\section{Conclusions}

In this paper we have derived the quantum limit on position
measurement of an oscillator by a generic linear response
detector, and on the noise temperature of a generic linear
amplifier.  The approach used makes clear what must be done to
reach the quantum limit; in particular, one needs a detector or
amplifier satisfying the ideal noise constraint of Eq.
(\ref{CorrQLCondition}), a demanding condition which is not
satisfied by most detectors.  Our treatment has emphasized both
the damping effects of the detector and its effective temperature
$T_{eff}$; in particular, we have found that the requirement of a
large detector power gain translates into a requirement of a large
detector effective temperature.

We thank Steve Girvin, Douglas Stone and especially Michel Devoret
for numerous useful conversations. This work was supported by the
W. M. Keck Foundation, and by the NSF under grants NSF-ITR 0325580
and DMR-0084501.
\begin{appendix}

\section{Derivation of Langevin Equation}

In this appendix, we prove that an oscillator weakly coupled to an
arbitrary out-of-equilibrium detector is described by the Langevin
equation given in Eq.~(\ref{Langevin}), an equation which
associates an effective temperature and damping kernel to the
detector; a similar perturbative approach for the problem of a
qubit coupled to a detector was considered by Shnirman et al. in
Ref. \onlinecite{ShnirmanSN}.

We start by defining the oscillator matrix Keldysh green function:
\begin{equation}
    \check{G}(t) = \left(
        \begin{array}{cc}
          G^K(t) & G^R(t) \\
          G^A(t) & 0 \\
        \end{array}
    \right)
\end{equation}
where $G^R(t-t') =  - i \theta(t-t') \langle
\left[x(t),x(t')\right] \rangle$, $G^A(t-t')  =   i \theta(t'-t)
\langle \left[x(t),x(t')\right] \rangle$, and    $G^K(t-t') = - i
\langle
        \{x(t),x(t')\} \rangle$.
At zero coupling to the detector ($A=0$), the oscillator is only
coupled to the equilibrium bath, and thus $\check{G}_0$ has the
standard equilibrium form:
\begin{equation}
    \check{G}_0(\omega) = \frac{\hbar}{m}
    \left(
        \begin{array}{cc}
            - 2 \textrm{Im } g_0(\omega)
         \coth \left(\frac{\hbar \omega}{ 2 k_B T_{bath}} \right) & g_0(\omega) \\
          g_0(\omega)^* & 0 \\
        \end{array}
    \right)
\end{equation}
where:
\begin{equation}
    g_0(\omega) = \frac{1}{\omega^2 - \Omega^2 + i \omega
    \gamma_0 / m}
\end{equation}
and where $\gamma_0$ is the intrinsic damping coefficient, and
$T_{bath}$ is the bath temperature.

We next treat the effects of the coupling to the detector in
perturbation theory.  Letting $\check{\Sigma}$ denote the
corresponding self-energy, the Dyson equation for $\check{G}$ has
the form:
\begin{equation}
\left[ \check{G}(\omega) \right]^{-1} =
    \left[ \check{G}_0(\omega) \right]^{-1} -
    \left(
        \begin{array}{cc}
          0 & \Sigma^A(\omega) \\
          \Sigma^R(\omega) & \Sigma^K(\omega) \\
        \end{array}
    \right)
    \label{Dyson}
\end{equation}
To lowest order in $A$, $\check{\Sigma}(\omega)$ is given by:
\begin{eqnarray}
    \check{\Sigma}(\omega) & = & A^2 \check{D}(\omega) \\
    & \equiv & \frac{A^2}{\hbar} \int dt \hspace{3pt} e^{i \omega t}
    \\
    && \left(
        \begin{array}{cc}
          0 & i \theta(-t) \langle [ \hF(t), \hF(0) ] \rangle \\
          -i \theta(t) \langle [ \hF(t), \hF(0) ] \rangle
            & -i  \langle \{ \hF(t), \hF(0) \} \rangle \\
        \end{array}
    \right) \nonumber
\end{eqnarray}
Using this lowest-order self energy, Eq.~(\ref{Dyson}) yields:
\begin{eqnarray}
    G^R(\omega) & = &  \frac{\hbar}
        { m(\omega^2 - \Omega^2) - A^2 \textrm{Re } D^R(\omega) + i \omega (\gamma_0 +
        \gamma(\omega)) } \nonumber \\
        \label{GR} \\
    G^A(\omega)  & = & \left[G^R(\omega) \right]^* \\
    G^K(\omega) & = & -2 i \textrm{Im } G^R(\omega) \times
        \nonumber \\
        && \frac{
        \gamma_0 \coth \left(\frac{\hbar \omega}{ 2 k_B T_{bath}} \right)
        + \gamma(\omega) \coth \left(\frac{\hbar \omega}{ 2 k_B T_{eff}}
        \right)}
        { \gamma_0 +
        \gamma(\omega)} \nonumber \\
        \label{GK}
\end{eqnarray}
where $\gamma(\omega)$ is given by Eq.~(\ref{gamma}), and
$T_{eff}(\omega)$ is defined by Eq.~(\ref{TEffDefn}).  The main
effect of the real part of the retarded $F$ Green function
$D^R(\omega)$ in Eq.~(\ref{GR}) is to renormalize the oscillator
frequency $\Omega$ and mass $m$; we simply incorporate these
shifts into the definition of $\Omega$ and $m$ in what follows.

If $T_{eff}(\omega)$ is frequency independent, then Eqs.
(\ref{GR}) - (\ref{GK}) for $\check{G}$ corresponds exactly to an
oscillator coupled to two equilibrium baths with damping kernels
$\gamma_0$ and $\gamma(\omega)$.  The correspondence to the
Langevin equation Eq.~(\ref{Langevin}) is then immediate.  In the
more general case where $T_{eff}(\omega)$ has a frequency
dependence, the correlators $G^R(\omega)$ and $G^K(\omega)$ are in
exact correspondence to what is found from the Langevin equation
Eq.~(\ref{Langevin}):  $G^K(\omega)$ corresponds to symmetrized
noise calculated from Eq.~(\ref{Langevin}), while $G^R(\omega)$
corresponds to the response coefficient of the oscillator
calculated from Eq.~(\ref{Langevin}).  This again proves the
validity of using the Langevin equation Eq.~(\ref{Langevin}) to
calculate the oscillator noise in the presence of the detector to
lowest order in $A$.

\section{Suppression of  Imaginary Parts of $\lambda$ and $\bS_{I Q}$}

Defining $    S_{I F}(\omega) = \int_{-\infty}^{\infty} dt \langle
I(t) F(0)    \rangle$, one has the relations:
\begin{eqnarray}
    \hbar \left(
        \lambda(\omega) - \lambda'(\omega)^*
        \right) &  = &
        -i \left[
            S_{I F}(\omega) - S_{I F}(-\omega)
    \right] \\
    \bS_{I F}(\omega) & = & \frac{1}{2} \left[
            S_{I F}(\omega) + S_{I F}(-\omega)^{*}
    \right]
\end{eqnarray}
which follow directly from the definitions of $\bS_{IF}$ and
$\lambda$.  Assuming now that we have a quantum limited detector
(i.e.~ Eq.~(\ref{CorrQLCondition}) is satisfied), a vanishing
reverse gain ($\lambda'=0$), and $k_B T_{eff} \gg \hbar \omega$,
we can use Eqs. (\ref{TEffDefn}) and (\ref{PropCond}) in
conjunction with the above equations to show:
\begin{eqnarray}
    \hbar \lambda(\omega) & = & - \gamma(\omega) \left[
        4 k_B T_{eff} \textrm{Im }\alpha + i 2 \hbar \Omega
        \textrm{Re } \alpha \right] \\
    \bS_{I F}(\omega) & = & \gamma(\omega) \left[
        2 k_B T_{eff} \textrm{Re }\alpha - i \hbar \Omega
        \textrm{Im } \alpha \right]
\end{eqnarray}
We see immediately that the imaginary parts of $\lambda$ and
$\bS_{IF}$ are suppressed compared to the corresponding real parts
by a small factor $\hbar \Omega / 2 k_B T_{eff}$.

\end{appendix}


\begin{thebibliography}{99}
\bibitem{TNNote}
    Note that our definition of the noise temperature
    $T_N$ conforms with that of Devoret and
    Schoelkopf \cite{DevoretNature}, but is slightly different
    than that of Caves \cite{Caves}.  Caves assumes the source is
    initially at zero temperature, and consequently
    uses the full quantum expression for its equilibrium noise.
    In contrast, we have assumed that $k_B T_S \gg \hbar \omega$.
    The different definition used by Caves leads to the result
    $k_B T_N \geq \hbar \omega / (\ln 3)$ as opposed to our Eq.~
    (\ref{TNQL}).
\end{thebibliography}
\end{document}